\RequirePackage{fix-cm}
\documentclass[twocolumn]{svjour3}          
\smartqed  
\usepackage{graphicx}
\usepackage{amsmath}
\usepackage{amssymb}
\usepackage{footnote}
\usepackage{subfigure}
\usepackage{url}
\usepackage{times}
\usepackage[numbers]{natbib}
\usepackage{soul}

\usepackage{mathptmx}      
\usepackage{latexsym}
 \journalname{Autonomous Robots}
\begin{document}

\title{Cooperative Queuing Policies for Effective Human-Multi-Robot
Interaction
}

\titlerunning{Effective Human-Multi-Robot
Interaction}        

\author{Masoume M. Raeissi        \and
        Alessandro Farinelli 
}

\authorrunning{M. Raeissi \and A. Farinelli} 

\institute{University of Verona \at
              Verona, Italy\\
              \email{masoume.raeissi@univr.it}           
           University of Verona\at
              Verona, Italy\\
              \email{alessandro.farinelli@univr.it}
}

\date{Received: date / Accepted: date}

\maketitle

\begin{abstract}
We consider multi-robot applications, where a team of robots can ask for the intervention of a human operator to handle difficult situations. 
As the number of requests grows, team members will have to wait for the operator attention, hence the operator becomes a bottleneck for the system. 
Our aim in this context is to make the robots learn cooperative strategies to decrease the time spent waiting for the operator.
In particular, we consider a queuing model where robots decide whether or not to join the queue and use multi-robot learning to estimate the best cooperative policy. 
In more detail, we formalize the problem as Decentralized Markov Decision Process and provide a suitable state representation, so to apply an independent learners approach. We evaluate the proposed method in a robotic water monitoring simulation and empirically show that our approach can significantly improve the team performance, while being computationally tractable. 

\end{abstract}

\section{Introduction}
\label{intro}
In many multi-robot scenarios, such as environmental monitoring \cite{Valada14} or search and rescue \cite{Wang07,ROB07}, one or few operators are required to interact with a team of robots to perform complex tasks in challenging environments. Robots, specially at field sites, are often subject to unexpected events, that can not be managed without the intervention of operators. For example, in an environmental monitoring application, robots might face extreme environmental events (e.g., water currents) or moving obstacles (e.g., animal approaching the robots). In such scenarios, the operator often needs to interrupt the activities of individual team members to deal with particular situations. 

The operator's monitoring and supervisory role in these scenarios becomes critical, particularly when the team size grows larger. To decrease the operator's monitoring task and give him/her more time to focus on robots that need attention, several approaches consider the concept of self-reflection \cite{scheutz07}, where robots are able to identify their potential issues and ask for the intervention of the operator by sending a request. However, large teams can easily overwhelm the operator with several requests, hence hindering the team performance. Consequently, team members have to wait for the operator's attention, and the operator becomes a bottleneck for the system. Queuing is a natural way to manage and address this problem. Previous research try to enhance the performance of the system (i.e., decreasing the time spent by robots waiting for the operator) considering various queue disciplines (e.g. First in First Out (FIFO) and Shortest Job First (SJF)) \cite{Lewis14,Chien12} or prioritizing such requests \cite{Rosenfeld15}. In both cases, the queue size may grow indefinitely as no robot will leave the queue before receiving the operator's attention. 

To deal with this problem, we focus on balking queue model \cite{naor69}, in which the users/agents (i.e robots requesting attention) can decide either to join the queue or balk. Such decisions are typically based on a threshold value, that is computed by assigning a generic reward for receiving the service and a cost for waiting in the queue to each agent. When applying this model to a robotic application, there is no clear indication how such thresholds can be computed. More important, this model does not consider the cost of balking (i.e. the cost of a potential failure that robots can suffer by trying to overcome difficult situations without human intervention). 
\begin{figure}
\centering
\includegraphics[width=0.95\columnwidth]{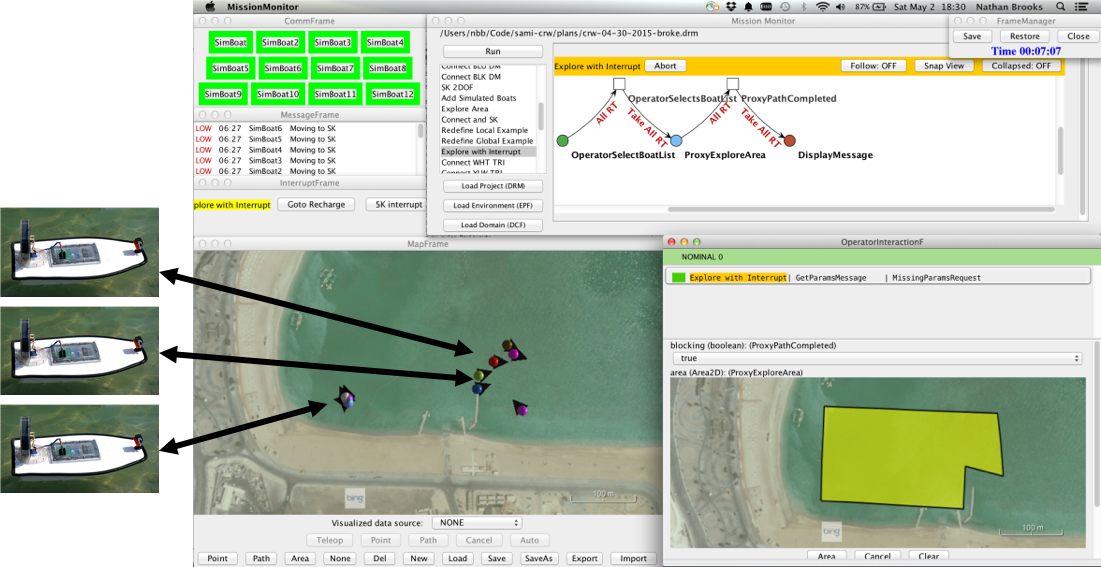}
\caption{Water monitoring simulation tool.}
\label{fig:airboat-architecture}
\end{figure}
Considering this, our focus is to devise an approach that allows the team of robots to learn cooperative balking strategies to make better use of a shared queue. Therefore, we frame the above problem as a Decentralized Markov Decision Process (Dec-MDP) in which, the team of robots must cooperate to optimize the team idle time. Finding optimal decentralized policies is often hard due to the partial observability and limited communications. Thus, our goal is to provide a scalable state representation by adding the state of the queue as an extra feature to the robots' local states and solve the underlying Dec-MDP problem using multiple independent learners \footnote{Some part of this work appears in \cite{Raeissi18}. That work describes basic ideas and preliminary results, here we provide a more detailed description of the methodologies, and more extensive empirical analysis.}. We illustrate that, this additional feature will improve the team performance over our main evaluation metric (i.e. team idle time). In more detail, this paper provides the following contributions to the state of the art: (i) We model the human-multi-robot interaction as a balking queue, in which the robots identify their needs and decide whether to interrupt the human operator or not. (ii) We formalize the problem as Dec-MDP and provide a tractable state representation to learn the balking policies for each robot. (iii) Finally, we evaluate the performance of our model by comparing the team idle time to the state-of-the-art queue disciplines. Overall, the experimental results show that, the use of our model decreases the total idle time upto 68\% over FIFO (without balking) and increases the team reward upto 56\% comparing to the other learning models.   

\section{Background}
\label{sec:1}
In this section, first we review the state of the art in robotic studies, where robots ask for operator's attention. Afterwards, we present a brief introduction to the Balking Queue\cite{naor69} model in which, users/agents decide to join the queue or not. Then, we present a brief review of Dec-MDP\cite{bernstein02,Goldman03} as the basis model for decision making under uncertainty in multi-robot scenarios. 

\subsection{Human-Multi-Robot Interaction}
\label{sec:2}
Human-multi robot interaction is an active field of research with many different research issues, including team plan representation \cite{Tambe97,Kaminka05}, multi-modal interaction \cite{Stoica13,Gromov16} and mixed initiative planning \cite{Bevacqua15}. Here, we consider approaches, which focus on how to allocate operator attention to a set of robots. Many work in this area consider that, the robots can perceive their situations and inform or ask the operator for help. For example, the work by \cite{Rosenthal10} proposes the idea of a single service robot asking for the help of humans. 

Considering larger multi-robot teams with small number of operators, several robots may need the operator's attention at the same time. Hence, the requests must be queued for being processed later on. Authors in ~\cite{Chien12,Lewis14} explore different queue disciplines to enhance the performance of the system. However, keeping robots idle until the operator becomes available might decrease the overall team efficiency. In contrast, we focus on a specific queuing model with balking property \cite{naor69}, where robots decide either to join the queue or not.

The concept of Adjustable Autonomy or mixed initiative has been the basis of many research in the field of human-multi-robot interaction. The key issue in this setting is to devise effective techniques to decide whether a transfer of control should occur and when this should happen. Different techniques have been proposed to address this challeng, for example, \cite{Horvitz99,collins00,barber00} consider that the robot will ask for human help/intervention when the expected utility of doing so is higher than performing the task autonomously, or when the uncertainty of the autonomous decision is high \cite{Gunderson99,Chernova09} or when the autonomous decision can cause significant harm \cite{Dorais98}. However, these decision making solutions usually have been considered as individual one-step decisions without considering the long-term cost or the consequences of decisions on other team members (if any). Our work also belongs to the this category, where autonomous robots in a multi-robot scenario can decide whether to wait for the operator or not. In particular, we focus on a specific queuing model with balking property \cite{naor69}. However, our focus is on finding cooperative strategies, where all robots learn concurrently to optimize the idle time of the system.

\subsection{Balking Queue Model}
\label{sec:3}
The first mathematical model of a queuing system with rational users was formulated by Naor \cite{naor69}. In his model, users upon their arrival decide according to a threshold value whether to join the queue or not (balk). The individual's optimizing strategy is straightforward, a customer will join the queue while $n$ other customers are already in the system if

\begin{equation}
R - n\cdot C \frac{1}{\mu} \geq 0
\label{B-Thresh}
\end{equation}
 
where a uniform cost $C$ for staying in the queue and a similar reward $R$ for receiving service are assigned to each user and $\mu$ is the intensity parameter of exponentially distributed service time. Thus, $n = \lfloor \frac{R \mu}{C} \rfloor$ serves as a threshold value for balking, that is if the number of users waiting in the queue is greater than $n$, the newly arrived user will not join the queue. In a multi-robot application, this threshold and decision must be computed carefully. Our focus is on showing how the elements (i.e. reward and cost) of balking strategy should be adjusted according to a practical robotics scenario. 

\subsection{Decentralized Markov Decision Process (Dec-MDP)}
\label{sec:4}
The decision of whether to join the queue or not for each situation of each robot will impact the future decisions of the entire team. As a result, we are concerned here with team sequential decision making problems, in which the team's utility depends on a sequence of decisions.

A Dec-MDP is defined by a tuple $\langle S, A, P, R \rangle$ where: $S$ is the set of world states which is factored into $n$ components, $S=S_1 \times ... \times S_n$. In a special case (i.e. Factored n-agent Dec-MDP), $S_i$ refers to the local state of agent $i$. In Dec-MDP, the state is jointly fully observable which means that the aggregated observations made by all agents determines the global state. $A=\times_i A_i$ is the set of joint actions, where $A_i$ is the set of actions for agent $i$. $P = S \times A \times S \rightarrow [0,1] $ is the state transition probability. $R = S \times A \rightarrow \mathbb{R}$ is the immediate reward. The complexity of Dec-MDP is nondeterministic exponential (NEXP) hard \cite{bernstein02}, hence learning is crucial. 
%
%
\section{Problem Formulation}
\label{sec:5}
We consider a water monitoring scenario, where several autonomous surface vessels are supervised by a human operator (see Fig. \ref{fig:airboat-architecture}). A set of events can happen to the platforms, such events may affect the normal behavior of the platforms and hinder their performance. Each event has a different probability of failure (see Table \ref{tab:1}), where requests with the higher probability of failure are more crucial to receive the operator's attention

Following previous works \cite{Chien12,Lewis14}, a central queue is provided to both the operator and the boats, where the operator can select one request at a time (i.e., in FIFO order) and assigns a specific sub-mission to resolve that request. 
A sub-mission is a plan specific recovery procedure, and this often requires a human interaction (i.e., the human directly selects which platforms should execute the interrupt sub-mission). We used three sub-missions, one for each class of requests, including: (i) sending a boat to the closest station to change/charge its battery. (ii) allowing/not-allowing a boat to go further (to the area that it might loose connection), and (iii) teleoperating a boat for traversing a specific area. 

\begin{table}
\caption{Different event types used in the experiments.}
\label{tab:1}       
\begin{tabular}{ll}
\hline\noalign{\smallskip}
Event Type ($E_j$) &  Prob. of Fail  \\
\noalign{\smallskip}\hline\noalign{\smallskip}
Battery-Recharge ($E_1$) & 0.9  \\
Traversing-Dangerous-Area ($E_2$) & 0.4  \\
Losing-Connection ($E_3$) & 0.2\\
\noalign{\smallskip}\hline
\end{tabular}
\end{table}

We assume that, whenever an event happens, the platform can detect the event. For example, a robot can perceive that its battery level is in a critical state, it must then decide whether to join the queue (i.e. sending the request and waiting for the operator) or balk (i.e., not sending the request) \footnote{While this may be a significant challenge in some domains, this is not the focus of our work.}.
The consequences or costs of balking are problem specific. In our model, when a failure happens, the operator should spend more time to fix the problem, hence failure as a result of balking, increases the idle time of the system. The goal is to minimize the time spent in the queue. 

Our proposal is then to train the robots in a stationary environment (i.e., stationary distribution functions with fixed arrival rate and service time), so that the robots can learn appropriate balking policies. Then, by applying the learned policies in similar scenarios, they will be able to optimize the team objective. More specifically, we consider the following model in our domain: 
The state space $S =  S_1 \times S_2 \times ... \times S_n$. $n$ is the number of boats. 
The local state of each boat $S_i$ is a tuple $\langle S_b , N_{tasks} \rangle$. $N_{tasks}$ shows the number of remaining tasks of boat $i$. In this application domain, each task is a location that should be visited by a specific boat. $S_b$ is the current internal state of boat $i$. More specifically $S_b \in \{E_j,\textbf{W}aiting,\textbf{F}ailed,\textbf{A}utonomy\}$, where $E_j$ refers to on of the request/event type in Table \ref{tab:1}. For example, the state tuple of a boat when it has 3 tasks to finish and the event \textit{Battery Recharge} occurs, would be $s = \langle E_1,3 \rangle$. $A_i$ is the set of actions for boat $i$ where $A_i \in \{ Join,Balk \}$. 
The reward function is designed to decrease the idle time (i.e. the time spent waiting for the operator). 

In general, there are two major approaches for learning in multi-robot scenarios \cite{panait05}. The first approach is called team learning and uses a single learner to learn the behavior for the entire team. In contrast, the second approach uses multiple concurrent learners, usually one for each robot, where each learner tries to learn its behavior. Each of these methods has its own advantages and disadvantages which make it preferable in different domains \cite{xuan02,panait05}. In particular, the major problems with team learning approach are the explosion of the state space (i.e., it keeps the states of the entire team), and the centralization of the learning approach that needs to access the states of all team members. 
Using the team learner in our application, the state space will be very large which decelerates the convergence to the optimal value. For example, for 5 boats with the above state representation, the state space will include more than one million states, hence requiring a prohibitive long time to estimate the optimal strategies for each state and action permutations. 
The main advantage of independent learners in our domain is that, this domain can be decomposed into subproblems (e.g. each boat holds its own state space) and each subproblem can be solved by one boat. In general, two main challenges arise in concurrent learning: credit assignment and non-stationary dynamics of the environment \cite{panait05}.
However, our application scenario has some special properties, that can be exploited to design a tractable model. First, the action selection at each step (i.e. when an event happens) only requires one agent to select either to join or balk. Hence, the reward can go directly to that agent. It is different from the situations, where all agents should decide at each step (i.e. joint actions), which results in the well-known credit assignment issue. 
However, when each boat considers only its local state without knowing the state of the queue, finding the optimal behavior for the team may become impossible, or the model may compute lower quality solutions. Therefore, we add the state of the queue to the local state of each boat, and then we use independent learners approach. 
To sum up, we consider three possible models: \\
\textbf{Team Learner (TL)}: a team learner has access to the joint state of all robots which is $S =  S_1 \times S_2 \times ... \times S_n$. When an event happens to a boat, the action $\langle Join,Balk \rangle$ for the corresponding boat will be selected and the state of the system will be updated. The update will only change the part of the state related to the corresponding boat. The Q-value of the team learner will be updated accordingly.\\
\textbf{Independent Learners - Unobservable Queue (IL-U)}: an independent learner) is used for each boat. Each boat observes only its local state $S_i = \langle S_b, N_{tasks} \rangle$. In this model, each boat updates its local Q-values interacting to the system and receiving the reward. \\
\textbf{Independent Learners - Observable Queue (IL-O)}: in this model, each boat in addition to observing its local state, has access to the size of the queue. The queue size shows the number of waiting boats inside the queue. The state representation of each boat in this model is:  $S_i = \langle S_b,N_{tasks},S_q \rangle$.\\
These models are different in their state representation, while the reward structure is the same for all of them: \\
(i) $R(S_t = S_i , A_t = Join) = R_S - (N_{q} \bar{\mu} + t_{serv})$. \\
(ii) $R(S_t = S_i , A_t = Balk) = R_F(\frac {\bar{\mu}}{\bar{\lambda}})+N_{q}$; 	\textit{if $S_{t+1} = F$}.\\ 
(iii) $R(S_t = S_i , A_t = Balk) = R_T$; 	\textit{if $S_{t+1} = A$}. 

$\bar{\mu}$ and $\bar{\lambda}$ are average service time and events arrival rate respectively. $N_{q}$ is the number of boats waiting, and $t_{serv}$ is the average time needed to resolve the request. $R_S = 1$, $R_F = -2$ and $R_T = 0.3$ are application specific parameters that must be tuned empirically. 

Finally, we use Q-Learning as the basis learning approach, while the same reward structure, same distribution functions for generating events and same distribution function for the service time are used for all models. We use Q-Learning because of its simplicity and good real time performance. Moreover, our goal is to propose the use of reinforcement learning in this novel context and not to provide a novel learning algorithm. In Q-Learning, each learner interacts with the environment (i.e. selects an action), receives the immediate reward and updates its state-action values (i.e. Q-values) as Eq.\ref{Q-Update}:
\begin{equation}
\resizebox{.91\hsize}{!}{$Q_i(s_i,a_i) \leftarrow Q_i(s_i,a_i) + \alpha(r_i + \gamma \max_{a^ {\prime} \in A_i} Q_i(s^ {\prime},a^ {\prime}) - Q_i(s_i,a_i)) $}
\label{Q-Update}
\end{equation}

where $r_i$ and $s^{\prime}$ are respectively the reward and the state observed by robot $i$ after performing action $a_i$ in state $s_i$; $a^{\prime}$ is the action in state $s^{\prime}$ that maximizes the future expected rewards; $\alpha$ is the learning rate and $\gamma$ is the discount factor.
Notice that, for Team Learner, there is only one Q-table (i.e., table of state-action values) to be updated, where, $s_i$ refers to full state representation (i.e. state of all robots) and $a_i$ is the action selected by learner for the specific robot with a request. This action will either add the event to the queue or not. As mentioned before, by exploiting specific feature of our domain, we can use an independent learner approach that is tractable and scalable. 

In the simulation, $\bar{\lambda}$ and $\bar{\mu}$ in the reward function are the same as $\lambda$ and $\mu$ for generating the requests and service. However, for a field deployment, these two parameters should be estimated by considering the average number of events being generated during an interval as $lambda$ (e.g. 20 events have been generated in 1 hour (or 60 minutes)), and the average time spent to fix each request as $\frac{1}{\mu}$ (e.g. 5 minutes to fix each request, hence 12 events can be fixed in 60 minutes).

The state of the queue, $S_q$, can be modified by robots' action (joining the queue) and the operator's action (leaving the queue). However, under the reasonable assumption that an arrival and a departure cannot happen exactly at the same time, only one entity can change the value of $S_q$ at a time. Moreover, the possibility of having more than one event at the exact same time is very low. In particular, for our scenario, it is safe to assume that, the time to change the state of the queue, is much lower than the time for a new event arrival. Under this assumption, even if two events happen within a short time interval, the first one will affect the state of the queue before the second arrives, hence the other robots will base their decisions on the updated queue size.

\begin{figure}
\centering
\includegraphics[width=\columnwidth]{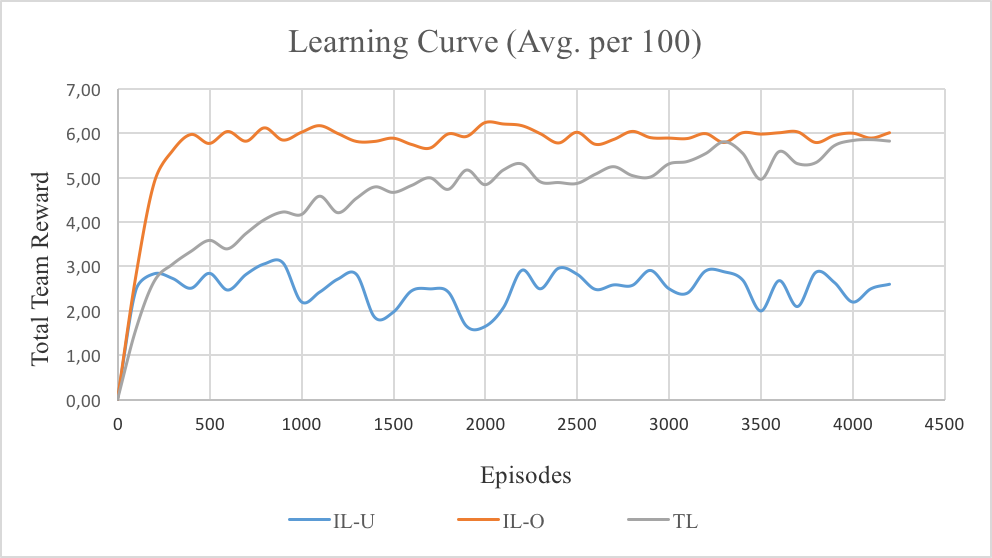}
\caption{Team accumulated reward in each episode of the learning phase (better viewed in color).}
\label{curve}
\end{figure}

\begin{figure}
\centering
\includegraphics[width=\columnwidth]{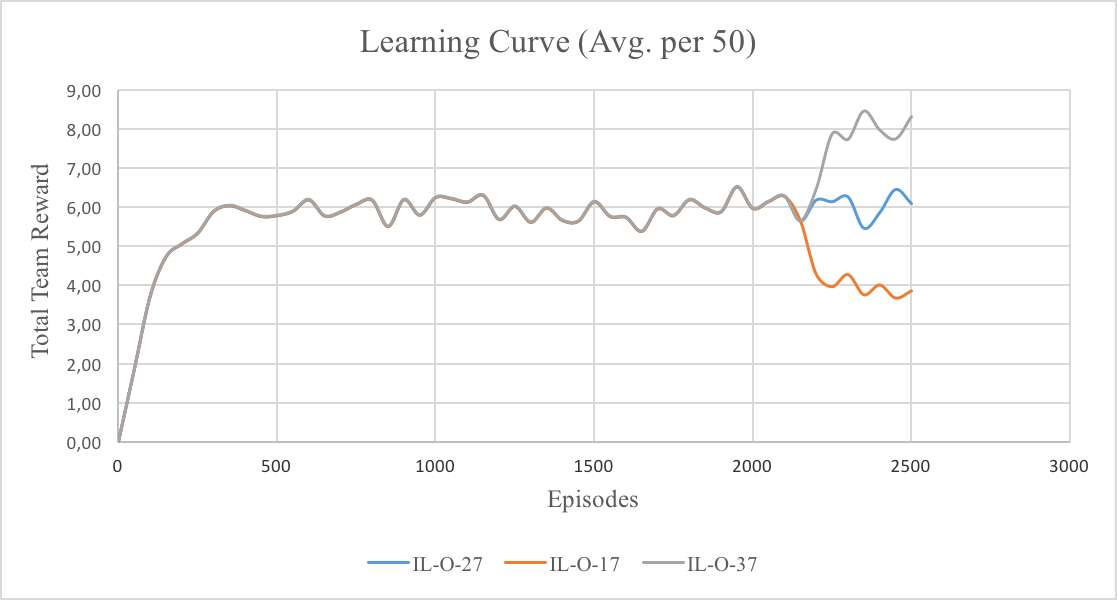}
\caption{Team accumulated reward in each episode of the learning phase (better viewed in color). From episode 2150, the service rate has been changed from 0.27 to 0.37 and 0.17.}
\label{incMu}
\end{figure}

\begin{figure}
\centering
\includegraphics[width=\columnwidth]{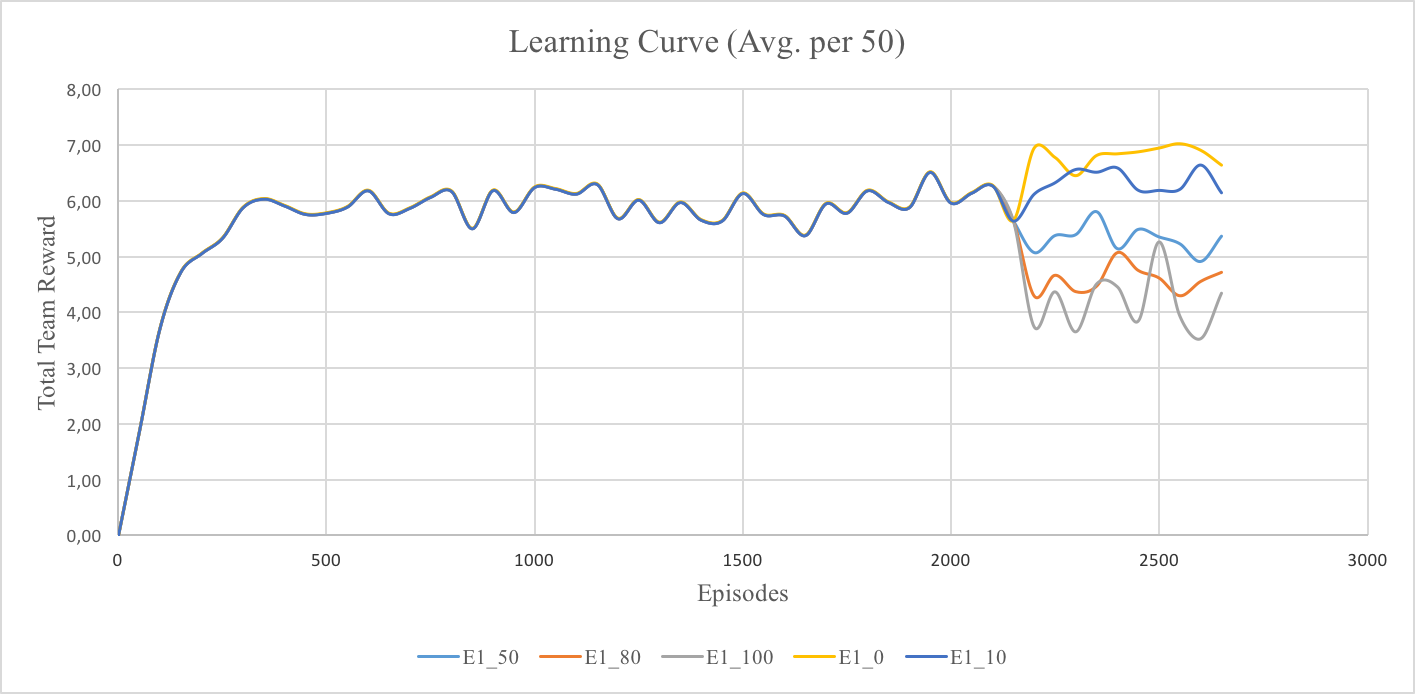}
\caption{Team accumulated reward in each episode of the learning phase (better viewed in color). After episode 2150, we vary the rate of each event type.}
\label{vary-lam}
\end{figure}

\section{Experimental Evaluation}
\label{sec:6}

\subsection{Learning Phase}
The learning phase of balking models starts by defining a list of locations (i.e., to be visited), and assigning those locations to boats. We consider 30 locations and 5 boats. Events, as in table \ref{tab:1}, will be generated within an exponential distribution with parameter $\lambda = 0.25$. The operator's speed, for resolving a request is selected from an exponential distribution with parameter $\mu = 0.27$. An episode (i.e., a run of the algorithm beginning from a start state to a final state) ends after the system encounters 20 events. For action selection in our model, we use $\epsilon\_greedy$ method with parameter $\epsilon = 0.1$. Our algorithm uses the learning rate $\alpha = 0.1$ and discount factor $\gamma = 0.9$ throughout the experiments, which are tuned empirically. Each episode of the learning phase starts with all boats in their Autonomy state (i.e. they do not need the attention of the human operator), then with arrival rate $\lambda$ an event may happen to one boat. We used a realistic estimation for parameter $\lambda$ and $\mu$ based on some experience on the total mission time, number of boats and number of locations. These numbers define well the type of scenarios we are interested in, where boats can operate most of the time in autonomy, but frequently need user's intervention. 

Fig. \ref{curve} shows the team rewards of each model, \emph{TL}, \emph{IL-U} and \emph{IL-O}, during the learning phase. The oscillation in the reward is due to the fact that, the robots learn their policies by trying new potentially sub-optimal actions. The training time (the sum over 4000+ episodes) for each model was about 88-95 hours. As we expected, the convergence rate of \emph{IL-O} is much faster than the \emph{TL}, while they both reach a similar team reward. This is due to the larger state space of \emph{TL} which needs more iterations to estimate the value for each state and action. Results also clearly show the importance of having access to the state of the queue to make better decisions. Since, the reward given to each action is related to the parameters $\lambda$ and $\mu$, we expect our policy to be dependent on these two parameters. Fig. \ref{incMu} shows how \emph{IL-O} adapts to changes in $\mu$, where we increase and decrease its value by 40\% during the learning phase. A sudden rise and drop happen respectively for each value, but then the system converges to a stationary state. Fig. \ref{vary-lam} shows another experiment, where we vary the ratio of event types during the learning phase. The events were generated with a uniform distribution up to episode 2150. After that, we vary the percentage of events from type 1($E_1$), which has the higher probability of failure (see Table \ref{tab:1}), while the rest of the events are uniformly distributed among $E_2$ and $E_3$. In more detail, we consider $E_1$-100, $E_1$-80, $E_1$-50, $E_1$-10 and $E_1$-0, where each shows the rate of $E_1$. Fig. \ref{vary-lam} illustrates that, as there are more $E_1$, the system will gain less reward due to the increasing rate of failures. After several iterations, the learning curve becomes stationary.  

\begin{figure*}
\centering
	\subfigure[Total team reward]{\label{reward}\includegraphics[width=0.33\linewidth]{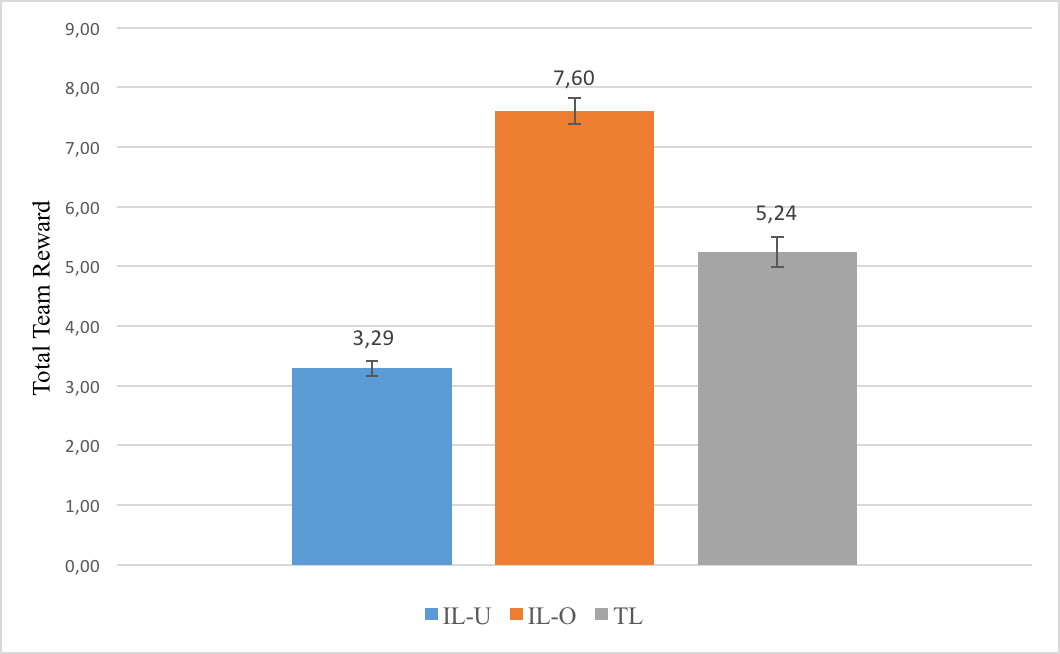}}
	\subfigure[Total idle time]{\label{waiting}\includegraphics[width=0.33\linewidth]{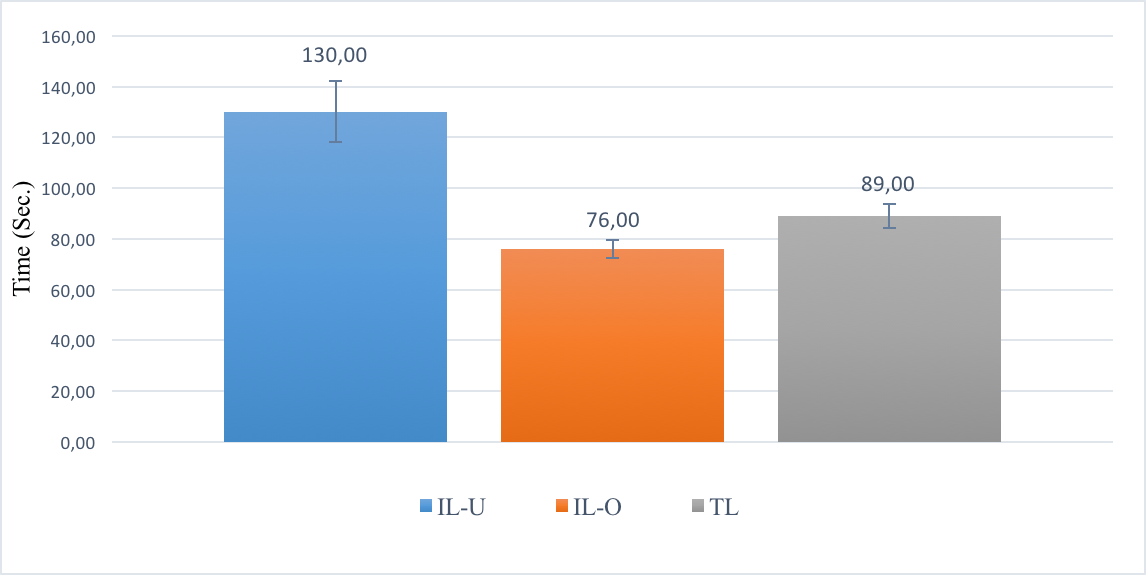}}
	\subfigure[Total idle time]{\label{all-queues}\includegraphics[width=0.33\linewidth]{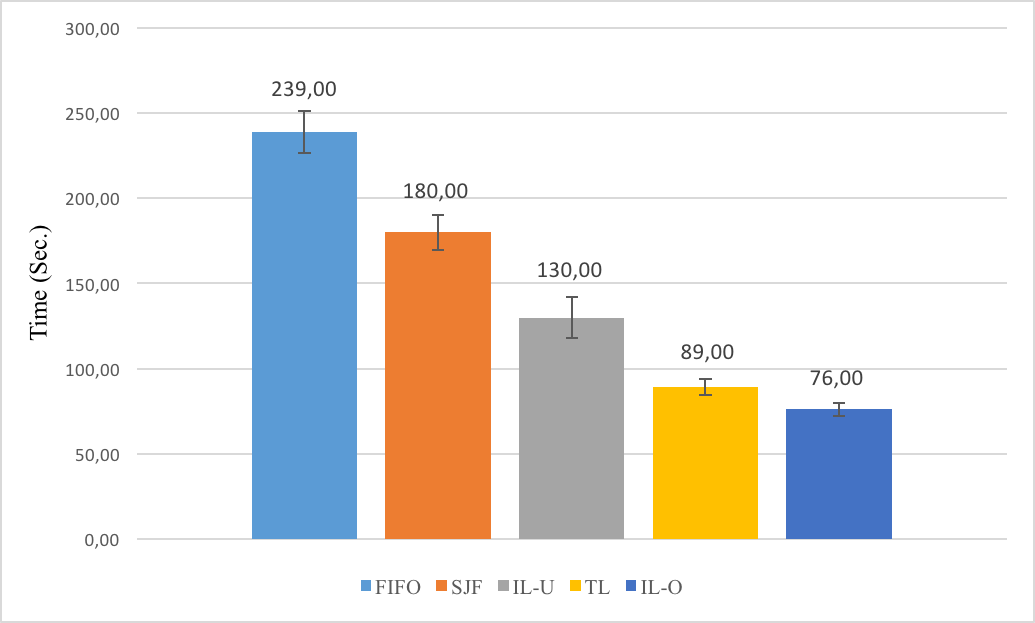}}
\caption{(a) and (b) show the team performance (together with the standard error of the means) for three learning models, while (c) compares balking models to non-balking models.}
\label{temp}
\end{figure*}

\subsection{Test Phase}
After the learning phase, we run 30 simulation executing the policy learned previously. In this first experiment, we use the same values for $\lambda$ and $\mu$ as used during the learning phase. Fig. \ref{reward} demonstrates the team reward for each learning models. A comparison on team reward between \emph{IL-O} and \emph{IL-U}, shows 56\% gain for \emph{IL-O}. Besides, a significant decrease (i.e. 40\%) on average waiting time is shown in Fig. \ref{waiting} when using \emph{IL-O} rather than \emph{IL-U}. One might expect the same reward value and idle time for \emph{IL-O} and \emph{TL}. However, the results on Fig. \ref{reward} and \ref{waiting} show better performance values for \emph{IL-O} than \emph{TL}. This difference is due to the fact that, \emph{IL-O} model keeps only the size of the queue or $S_q$ (i.e. it does not consider which boats are waiting in the queue), while \emph{TL} maintains the state of all boats which are in their $\textbf{W}aiting$ state (i.e. $S_b = \textbf{W}$). For example, whenever two boats waiting in the queue (assuming the other features, e.g., severity are the same), \emph{IL-O} will map the state to $S_q = 2$, while \emph{TL} will differentiate the states depending on which two boats are inside the queue. Since, the boats are homogeneous in our domain, \emph{IL-O} results in better performance by abstracting away features that do not have a significant impact on the reward. This also makes \emph{TL} to converge slower than \emph{IL-O}, due to the larger state space of \emph{TL} which needs more iterations to estimate the value for each state and action.

\begin{figure}
\centering
\includegraphics[width=\columnwidth]{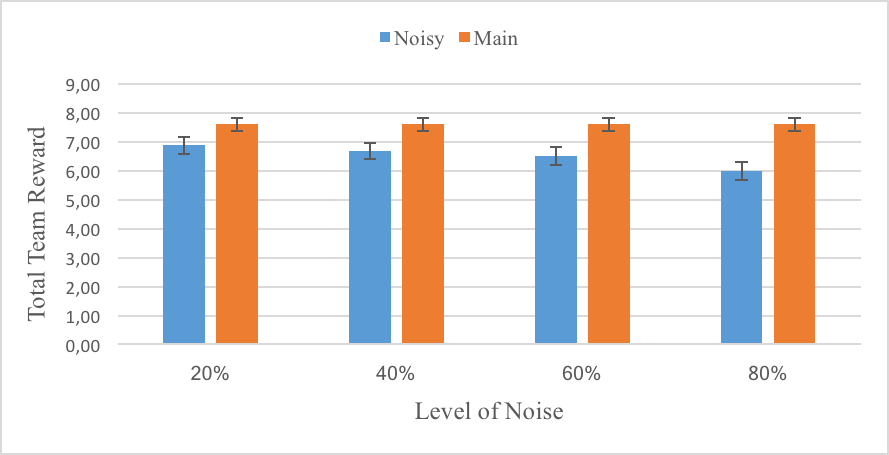}
\caption{Total team reward (together with the standard error of the means) for IL-O (main) with different levels of noise on $\lambda$ and $\mu$ (noisy).}
\label{mu-lambda-stdev-rew}
\end{figure}

\begin{figure}
\centering
\includegraphics[width=\columnwidth]{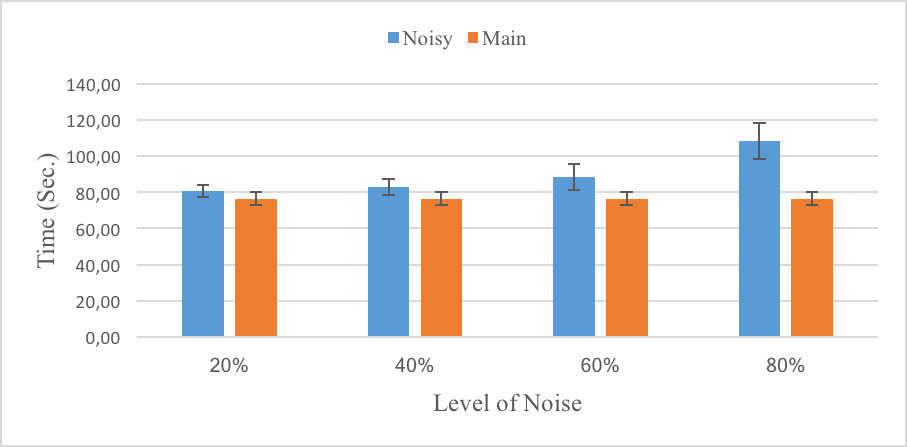}
\caption{Total idle time (together with the standard error of the means) for IL-O (main) with different levels of noise on $\lambda$ and $\mu$ (noisy).}
\label{mu-lambda-stdev-time}
\end{figure}
Next, we compare the behavior of queues with and without balking property (e.g., FIFO and SJF). For FIFO and SJF, we use the same event rate $\lambda$ and service rate $\mu$. In these two queuing models, boats always join the queue regardless of their request types and the queue size. Fig. \ref{all-queues} shows the team idle time for FIFO, SJF and three learning models. FIFO without balking, has the worst performance, since boats wait for the operator until he/she becomes available. In contrast, \emph{IL-O} approach outperforms all other models. In more detail, it decreases the time up to 68\% comparing to FIFO. In general, the results in Fig. \ref{all-queues} indicate that, using balking models significantly decreases the idle time of the team even though, some events may result in failures. This is acceptable in our domain, since the penalties for failures are not critical but only result in a finite increase of time. 

To validate the noise sensitivity of our proposed model \emph{IL-O}, we consider a set of experiments as follow.
We consider adding the same level of noise, according to a uniform distribution, to both parameters $\lambda$ and $\mu$ during the test phase. Fig. \ref{mu-lambda-stdev-rew} shows the team reward and Fig. \ref{mu-lambda-stdev-time} shows the team idle time for different levels of noise. The results show that, the approach is able to cope with a significant amount of noise on both $\lambda$ and $\mu$.  

\section{Conclusions}
In this paper, we propose the use of balking queue to model human-multi-robot interactions when the autonomy of robots allow them to decide whether to wait for the operator or not. We frame the problem as a Dec-MDP in which, each robot observes its local state and the state of the queue and cooperates with other agents to optimize the use of a shared queue. We apply independent Q-Learning to find these cooperative strategies in a water monitoring multi-robot simulation. We consider three different models (TL, IL-U, IL-O), and our results clearly show that an independent learner approach where the state of the queue is accessible to the platforms performs best. Furthermore, the empirical results related to the noisy estimation for $\lambda$ and $\mu$ (Fig. \ref{mu-lambda-stdev-rew} and \ref{mu-lambda-stdev-time}), suggest that the approach is able to cope with a significant amount of noise.


%
%


\bibliographystyle{spmpsci}      
\bibliography{auroma}   


\end{document}